# Ultrafast Chemical Imaging by Widefield Photothermal Sensing of Infrared Absorption

Yeran Bai,[1,2,3,4†] Delong Zhang,[3,4†] Lu Lan,[4,5] Yimin Huang, [4,6] Kerry Maize,[7] Ali Shakouri,[7,*] Ji-Xin Cheng[3,4,5,6,*]

[1]National Laboratory on High Power Laser and Physics, Shanghai 201800, China.

[2]Key Laboratory of High Power Laser and Physics, Shanghai Institute of Optics and Fine Mechanics, Chinese Academy of Sciences, Shanghai 201800, China.

[3]Department of Electrical & Computer Engineering, Boston University, Boston, MA 02215, USA.

[4]Photonics Center, Boston University, Boston, MA, 02215 USA.

[5]Department of Biomedical Engineering, Boston University, Boston, MA 02215, USA.

[6]Department of Chemistry, Boston University, Boston, MA, 02215 USA.

[7]Birck Nanotechnology Center, Purdue University, West Lafayette, IN 47906, USA.

[*]Corresponding author.  E-mail: jxcheng@bu.edu, shakouri@purdue.edu

[†] Equal contributions.

Infrared (IR) imaging has become a viable tool for visualizing various chemical bonds in a specimen. The performance, however, is limited in terms of spatial resolution and imaging speed. Here, instead of measuring the loss of the IR beam, we utilize a pulsed visible light for a high-throughput, widefield sensing of the transient photothermal effect induced by absorption of single mid-IR pulses. To extract such transient signals, we built a virtual lock-in camera synchronized to the visible probe and IR light pulses with precisely-controlled delays, allowing sub-microsecond temporal resolution determined by the probe pulse width. Our widefield photothermal sensing (WPS) microscope enabled chemical imaging at a speed up to 1250 frames per second, with high spectral fidelity, and offering sub-micron spatial resolution. With the capability of imaging living cells and nanometer-scale polymer films, WPS microscopy opens a new way for high-throughput characterization of biological and material specimens.

**Introduction**

Vibrational imaging methods offer a new window to characterize samples based on spectroscopic signatures of chemical bonds. Raman and infrared (IR) spectroscopy have long been used to interrogate materials by probing molecular vibrations without exogenous labels. Spontaneous Raman microscopy offers sub-micrometer spatial resolution imaging capability, but suffers from the low acquisition rates (*1, 2*). With the advent of coherent Raman scattering techniques, video-rate imaging speed has been demonstrated to characterize biological and pharmaceutical samples (*3, 4*). However, detection of the extremely small Raman cross sections ($10^{-30}$ cm$^2$sr$^{-1}$) limits the sensitivity. On the other hand, the IR absorption offers larger cross sections ($10^{-22}$ cm$^2$sr$^{-1}$) that enables adequate sensitivity. Fourier-transform IR (FTIR) spectrometer, together with its attenuated total reflection accessories, is the typical instrument of the technique and has been extensively employed in the fields ranging from polymer science, pharmaceuticals to biological research (*5-7*). Coupling focal plane array detectors to FTIR systems allows simultaneously acquiring spatially resolved spectra, greatly improving the throughput for characterization of inhomogeneous samples (*8*). Unlike the conventional FTIR instrument based on interferometry and low-brightness globar excitation, discrete IR spectroscopic imaging techniques utilize tunable quantum cascade laser (QCL) with much higher photon flux per wavenumber, which enables real-time IR imaging (*9*). However, the long incident wavelengths in the mid-IR region determines the spatial resolution at several to tens of micrometers, which is not sufficient to resolve microstructures such as in biological cells.

To address the resolution issue, near-field approach provides a way to surpass the fundamental limitations by combing atomic force microscopy (AFM) with IR spectroscopy (*10-12*), where the AFM cantilever changes the oscillation amplitude due to the surface thermal expansion induced by the absorption of the mid-IR light. The spectra at nanoscale localization are obtained by recording the amplitude change while sweeping the wavelengths of the mid-IR light source. With the capability of providing high spatial resolution chemical mapping, AFM-IR has been a valuable tool to study block copolymer system where the domain size is typically at tens of nanometers (*13*). This technique shares the inherent drawback of tip-based imaging modality of low acquisition speed. Additionally, although some work showed the capability of investigating samples in aqueous environment using the total internal reflection of an IR prism to minimize the influence of water, sophisticated setup and data processing procedure make it unsuitable for routine use (*14*, *15*).

In contrast, a non-contact probe, such as a visible laser, can reduce the limits on sample preparation and provide higher imaging speed. Recently, our group developed a mid-infrared photothermal (MIP) microscope using a visible laser to probe the IR absorption induced thermal lensing effect in the sample, providing chemical imaging capability with sub-micrometer resolution and depth resolution (*16*), which fills the gap between FTIR and AFM-IR microscopy. When the IR wavelength is tuned to the absorption peak of the sample, the co-propagated probe beam will change its divergence due to the thermal-induced local refractive index change. We demonstrated chemically-selective imaging of live cells and organisms (*16*, *17*). For non-transparent samples, a backward-detected photothermal microscopy was developed to allow chemical mapping of active pharmaceutical ingredients and excipients of drug tablets (*18*). In the meantime, optical probing of IR absorption has also been implemented by other groups. Furstenberg et al. demonstrated photothermal imaging of materials using a visible laser probe (*19*). Erramilli and co-workers investigated the nonlinear photothermal phenomena causing spectral peak splitting at different phase state of liquid crystal (*20*, *21*). The Sander group demonstrated photothermal IR imaging of a thin polymer film with spatial resolution of 3.1 μm (*22*) and reported photothermal IR imaging of mouse brain tissue slices targeting protein amide I band consistent with hematoxylin and eosin staining results (*23*). Recently, the Hartland group reported a counter-propagation scheme at 300-nm resolution and its application to characterize the perovskite-based solar cell (*24*, *25*). In these pixel-by-pixel scanning implementations, the imaging speed is limited by the pixel dwell time needed to cover the photothermal decay process. Alternatively, based on speckle pattern and Mie scattering, respectively, widefield photothermal imaging has been demonstrated (*26*, *27*). Typical acquisition time for a field of view of 17 × 17 μm$^2$ was around 20 seconds. Such speed is insufficient for study of living cells or high-through screening purpose.

Here, we demonstrate widefield photothermal sensing (WPS) microscopy that allows ultrafast chemical imaging at a speed up to 1250 frames per second. To enable high-throughput detection of IR absorption, a multi-element photodetector, such as a camera, is required. However, generic cameras are not fast enough to resolve the transient thermal process at the microsecond level, i.e. one million frames per second. To achieve high temporal resolution using regular cameras, time-gated detection using pulsed light was demonstrated for mapping electronic currents in integrated circuits (*28*, *29*). Here, we borrow this concept and build a virtual lock-in camera, where the frames are synchronized to the probe pulse and the IR pulse at the same repetition rate with precisely-controlled time delays. Our method enables time-resolved imaging of the transient thermal process using a regular camera, with a temporal resolution determined by the probe pulse width.

Furthermore, to enable efficient delivery of the IR laser to the sample and reflection of the probe photons to the camera, we adopt the widely available silicon wafers as substrate for its transparency in the IR window and high reflectance of visible photons. The silicon substrate further enhances the WPS signal by accelerating the heat dissipation. Silicon has high thermal conductivity (150 Wm$^{-1}$K$^{-1}$) compared to other IR transparent materials such as CaF$_2$ (10 Wm$^{-1}$K$^{-1}$), which avoids heat accumulation and allows faster imaging. Collectively, these innovations enabled ultrafast detection of IR-induced photothermal signals in a widefield manner.

**Results**

**WPS microscope**

The WPS imaging system is based on a widefield reflection microscope (**Fig. 1**, detailed in Materials and Methods). A pulsed blue light emitting diode (LED) was used to illuminate the sample through a 4-f lens system, a 50/50 beam splitter, and an objective. The light reflected from the sample was then collected by the same objective and beam splitter, and recorded by a camera via a tube lens. The pump source was provided by a nanosecond-pulsed mid-IR laser, which was weakly focused from below the sample, through the silicon wafer. A chopper was used to modulate the IR pulse train to accommodate the speed of the camera. The master clock of the system was provided by the IR laser, monitored by a mercury cadmium telluride (MCT) detector through a residual IR beam picked up from a CaF$_2$ plate.

**Widefield lock-in measurement of transient thermal signal**

To extract the transient WPS signal, we developed a virtual lock-in camera synchronized to the IR laser repetition rate with precise delays of the probe pulse. By utilizing a pulsed probe light, the temporal resolution of the system was determined by the pulse width of the probe light, which was around 900 ns. **Fig. 2a** shows the block diagram of widefield lock-in detection. The master clock from the MCT was used to trigger the function generator, which sent square wave triggers to the camera, chopper, and LED. The time delay of the probe pulse relative to the pump pulse was controlled electronically by the function generator. The camera provides a frequency division function that enables exposing under 2500 Hz with a 20 kHz external trigger frequency. Furthermore, the optical chopper frequency was locked to the camera exposure period to ensure complete block of the IR pulses when necessary.

The actual pulse trains of pump and probe pulses relative to the camera exposure were measured and plotted in **Fig. 2b-c**. The camera exposure indicator is plotted in channel 3 with high voltage level represents the actual camera exposure period. **Fig. 2c** shows a zoom-in image of the pump and probe delay. As a result, the IR pulses were chopped into bursts with as few as eight pulses during a camera exposure, recorded by the visible probe pulses with controlled delays. We define the images with mid-IR pulses as 'hot' frames, and those without as 'cold' frames. Therefore, the final data became an image stack with alternating hot and cold frames. The WPS signal was formed by accumulating the subtraction of adjacent 'hot' and 'cold' frames (**Fig. 2d**). Furthermore, the thermal decay profile can be mapped by scanning the probe delays.

**Temporal profile and chemical selectivity of WPS microscope**

To characterize the temporal resolution, we performed time-resolved WPS imaging of 486-nm thick poly(methyl methacrylate) (PMMA) film on silicon substrate. The IR pump was tuned to 1728 $cm^{-1}$, the C=O absorption peak in PMMA. The probe width was 914 ns and each image was acquired at the speed of 2 Hz. The pump and probe power at the sample were around 5.1 mW and 1.6 mW. By subtracting the cold frames by hot frames, i.e. contrast reversed for better visualization, WPS imaging at various delays were acquired (**Fig. 3a**). The bright spot in the center of field of view corresponded to the weakly focused IR spot. The signal intensity at the center of each image was used to plot the temporal profile of WPS signal, with time delay ranging from -0.97 μs to 3.89 μs (**Fig. 3b**). The video of the decay can be found in **Supplementary Video 1**. The data points were well fitted into an exponential decay $e^{-t/\tau}$, where $t$ was the time delay and $\tau = 1.1$ μs was the decay constant, indicating a fast cooling time. Note that the decay constants should not be confused with temporal resolution, which is determined by the probe pulse width, i.e. 914 ns, in WPS, limited by the LED.

We further demonstrated the spectral fidelity of our WPS microscope. We tuned the pump wavelengths at the time delay where the WPS signal was maximized. The raw spectrum was normalized by the pump power. The curve was the reference FTIR spectrum measured with a commercialized FTIR spectrometer (VERTEX 70v, Bruker). As shown in **Fig. 3c**, a good agreement was obtained between the WPS signal and the standard spectrum.

**Sub-micrometer spatial resolution**

In order to evaluate the spatial resolution, WPS imaging of polymer film patterns and beads were performed (**Fig. 4**). The 'MIP' letters were etched off a PMMA film around 310 nm thick using electron-beam lithography and the bare silicon showed higher reflectivity compared to the rest areas. The widefield reflection image is shown in **Fig. 4a**. The letter 'I' was 2 μm in width and 10 μm in length. WPS imaging of the pattern was acquired within 0.5 s with a cold frame minus a hot frame, showing clear contrast at the etching boundaries (**Fig. 4b**). The intensity profile along the line was extracted, as indicated in **Fig. 4b**. Nine adjacent lines were averaged to perform the first derivative (**Fig. 4c**). The theoretical value for optical resolution with 0.66 NA objective and 450 nm illumination is calculated to be 0.42 μm. The Gaussian fitting result showed a full-width-half-maximum (FWHM) of 0.51 µm, indicating a sub-micrometer resolution, which is consistent with the calculated value.

To test the detection limit of the WPS microscope, we measured the WPS signal at different thickness with fixed pump-probe delay. As a result, the minimal detectable thickness was around 159 nm, which is about an order smaller compared with a typical confocal Raman microscopy (*[30]*). Notably, the detection limit is comparable with the AFM-IR technique (*[31]*) and can be improved by coupling low-noise cameras.

To demonstrate the capability of imaging micro-particles, we performed WPS imaging of 1 µm PMMA beads. The reflection image of beads on silicon wafer is shown in **Fig. 4a**. Although the magnification of the system was not well optimized for such small particles, the images still has sufficient pixels for each bead. WPS imaging of the same area was acquired with the signal averaged for 0.5 s (**Fig. 4b**). Individual beads were clearly resolved at the PMMA C=O peak at 1728 cm$^{-1}$, while no contrast was shown at the off-resonance wavelength at 1808 cm$^{-1}$ (**Fig. 4e,f**). Here, for consistency, we used the same settings for WPS imaging. Note the field of view is only about a quarter of the previous "MIP" pattern in **Fig. 4a-c**, which potentially provides 4 times increase in imaging speed without any changes of the instruments.

**Ultrafast chemical imaging of a nanoscale film at the shot noise limit**

We demonstrated WPS imaging of a thin PMMA film (486.89 ± 0.16 nm) and evaluated the signal to noise ratio (SNR) as a function of imaging speed (**Fig. 5**). A total of 1054 frames was captured in 410 ms, i.e. 0.39 ms exposure time per frame, with a field of view of 136 µm by 108.8 µm. This results in a WPS imaging speed of 1250 frames per second. For noise measurement, a reference experiment was performed with the pump turned-off. Therefore, the subtracted results were pure noise from the camera and the probe photon fluctuation. The SNR was calculated from the center region (25 pixels) by the ratio of the mean difference between signal and reference to the standard deviation of the noise. **Fig. 5a** shows a single frame from the time trace, with the SNR calculated to be 2.3. Notably, the single frame chemical image was based on only eight IR pump pulses. To further improve SNR, frame averaging was used as a trade-off of imaging speed (**Fig. 5b-e**). We measured and calculated the SNR of each frame and fitted the results with a power function (**Fig. 5f**). The solid curve was the fitting result of different accumulating frames. The exponent term was 0.48, close to the theoretical limit of 0.5, indicating the setup has been optimized for working near shot-noise limit. A video with the imaging speed of 1250 frames per second is available (**Supplementary Video 2**). The signal intensities were relatively stable during the 0.42 seconds recording period, implying no sign of sample damage or photobleaching.

**Live cell imaging by WPS microscopy**

To demonstrate the capability of WPS microscopy for biological samples, we performed WPS imaging of living SKVO3 human ovarian cancer cells. **Fig. 6a** shows the reflection bright field image, showing a cell with lipid droplets around the nucleus. The WPS image at 1744 cm$^{-1}$ shows the distribution of lipid droplets, where individual droplets co-localized with the reflection image and no obvious signal showed up at the nuclei area (**Fig. 6b**). By tuning IR to 1656 cm$^{-1}$ of protein amide I band, the protein contents emerged to the contrasts in **Fig. 6c**. The proteins are more uniformly distributed in the cells compared to the lipid droplets. Note there is some residual absorption of water at this wavelength, which was not observed as no contrast was shown in the medium region outside the cell. It is consistent with previous study (*[16]*). This lack of contrast from water is due to the large heat capacity of water, resulting in a minimal change in temperature. Consequently, when tuned IR to the off-resonance wavelength at 1808 cm$^{-1}$, no contrasts were observed (**Fig. 6d**). For consistency in instrument, we used the same field of view for cell imaging, at a speed of 2 Hz. The final images were cropped to 40 μm by 40 μm for better visualization. Taken together, these data highlight the capability of imaging chemical components inside living cells with high speed.

**Discussion**

We demonstrated a WPS microscope that probes the thermal-induced reflectivity change with a camera through lock-in detection. The time-resolved mapping of heat dissipation was made possible by synchronization of frame capture with modulated IR pump and pulsed visible probe. Given the results of sub-microsecond temporal resolution, ultrafast chemical imaging speed (up to 1250 frames per second) and sub-micrometer spatial resolution, our method opens a new way to study highly dynamic processes with motion-blur-free observation. It also opens a door for reagent-free, high-throughput screening for fields ranging from pharmaceutical industry to cell biology (*32-34*). Another potential application is to differentiate the chemical and morphological features of a sample based on the signal level and decay speed.

There is still plenty of room to improve the performance of the WPS system. To further boost the imaging speed, a camera with high well depth pixels can be adopted. Because the WPS signal comes from small AC signals on top of a strong DC background, increasing the full well capacity of the image sensor helps to reduce the averaging time. The current camera provides a dynamic range of 58 dB with a full well capacity of 19 ke$^-$ and dark noise of 23 e$^-$. As an example, the commercialized camera (Q-2HFW-CXP, Adimec) has 2 million full well capacity and 63 dB dynamic range, which can be used to increase the speed by 10 times while maintaining the same SNR level. Furthermore, denoising methods (*35*) can be applied to further remove the noise in the X-Y-time data cube.

Currently, we are using counter-propagation of visible and IR beams. To broaden the applications, a co-propagation scheme can be developed, where the IR and visible beam travels towards the same direction, leaving space for thick samples. Previous work in backward-detected MIP (*18*) has demonstrated the implementation of co-propagation with the IR and visible beam sharing the objective. Alternatively, oblique illumination of IR beam can be used to separate the optical elements for IR and visible for better alignment.

Furthermore, it is of great importance to chemically image nano-sized particles in a biological or pharmaceutical environment. For this purpose, rigorous designs of the substrate and the embedding medium are needed. The bi-layered substrate of Si and $SiO_2$ structure has successfully been used in the interferometric reflectance imaging sensor to probe biomass accumulation and single virus (*36, 37*). Hence, by comparing the interference intensity at hot and cold states, biological nanoparticles with chemical specificity can be effectively mapped.

**Materials and Methods**

**WPS microscope**

The pump source was provided by a mid-IR optical parametric oscillator (Firefly-LW, M-Squared Lasers), tunable from 1175 to 1800 $cm^{-1}$ (8.51 μm to 5.56 μm), in the fingerprint region. A $CaF_2$ plate was used to pick off partial IR beam and sent into an MCT (PVM-10.6, Vigo System) detector. Since the camera shutter speed cannot catch each single pulse, we added an optical chopper (MC2000B, Thorlabs) to modulate the pump IR pulses. The chopper was working at 1250 Hz with the duty cycle of 50%. To reduce the rise and fall time of pulses at the edge of chopper blade, a gold-coated off-axis parabolic mirror (OAPM) with focal length of 101.6 mm was used to focus the pump beam at the blade. Such focal length was selected for the moderate focal spot size and ease of adjustment consider strong astigmatism if misaligned. Another OAPM with the same focal length was set up to collimate the IR beam. The modulated pump beam was guided by gold mirrors and then weakly focused on the sample using a $CaF_2$ meniscus lens. The visible probe beam was provided by a high-power LED working under pulsed operation mode (UHP-T-SR, Prizmatix). Due to the transient nature of the thermal diffusion process (*38*), we pulsed the LED output at sub-microsecond level to obtain the time-resolved signal. The central wavelength of the LED is 450 nm and the spectral width is 22.6 nm. The LED emitter was conjugated on the back focal plane of the objective (SLMPLN Olympus, 20x, NA 0.25, 440 Leica, 40x, NA 0.66) through a pair of 4-f lens and 50/50 plate beam splitter. The sample was illuminated by the parallel beam and the field of view was 136 μm by 108.8 μm for the 20x objective. The sample-reflected light was collected by the same objective and traveled back through the beam splitter to be imaged on a complementary metal oxide semiconductor (CMOS, IL5, Fastec Imaging) by a tube lens system. The camera shutter speed was set to 2500 frames per second at the abovementioned field of view.

**Film sample preparation**

Double-side polished silicon wafers with the thickness of 100 μm (University Wafer) were used as the substrates in film and pattern imaging. The silicon wafers were cleaned through multiple steps of solvent rinse, in the sequence of toluene, acetone, isopropanol and deionized water, followed with $O_2$ plasma (300 sccm $O_2$ flow rate, 300 W and 15 minutes) before the spinning coating process. To fabricate the PMMA films, PMMA 950 A4 solution (MicroChem) was spin-coated onto the cleaned silicon wafer, at a speed of 1000 rpm/s for 45s. After soft-baked at 180

°C for 5 minutes, the film thickness was measured by an ellipsometer (Rudolph AutoEL IV-NIR-II). To fabricate the pattern, PMMA 950 A4 solution was spin-coated onto the silicon wafer at a speed of 4000 rpm/s for 45s to get the electron beam resist layer. After soft-baked at 180 °C for 5 minutes, the "MIP" patterns were fabricated onto the PMMA layer with electron-beam lithography using a Zeiss Supra 40 VP SEM equipped with an e-beam blanker and through subsequent development with methyl isobutyl ketone/isopropanol solvent mix (1:3 in volume).

**Thermal decay characterization**

The pump-probe delay was pre-measured with an oscilloscope and noted as the reference for follow-up experiments. The delay was tuned using the function generator marked with phase shift in degrees. The entire 360° cycle corresponds to the pump pulse period of 50 μs.

**Live cell imaging**

The SKVO3 cells were cultured on a double polished silicon wafer for 24 hours. The cells were moved to microscope for immediate imaging after washed off cell culture medium and rinsed with PBS for 2 times.

**Data acquisition and image processing**

The images were captured at the camera shutter speed of 2500 Hz with the sensor resolution of 400 by 320 pixels. To maintain the image quality, 12-bit raw tiff image type was selected. For the model in the listed experiments, an acquisition process of 5 s would generate 12500 frames equals to 3.1 GB data size. We took the advantage of the onboard memory of the camera to store the huge amount of data and batch transfer the image stacks after the acquisition finishes. Therefore, we were not limited by the data transfer speed between the camera and the local memory drives. The downloaded images were analyzed by imageJ and custom code based on LabVIEW (National Instruments).

**Acknowledgments**

**General**: We thank Yuying Tan for the cell culture and Celalettin Yurdakul for helpful discussions.

**Funding:** This work was supported by a Keck Foundation Science and Engineering grant and R01 GM126049 to J.X.C. Y.B. acknowledges the UCAS (UCAS [2015]37) Joint Ph.D. Training Program for financial support.

**Author contributions:** Y.B., D.Z., and J.X.C. designed the experiments; D.Z. and K.Z. did the preliminary test; Y.B. and D.Z. performed the experiments and analyzed the data; Y.B., D.Z., L.L., and J.X.C. discussed the results; Y.H. prepared the film and pattern samples. D.Z., Y.B., and J.X.C. wrote the manuscript with the input from A.S. and K.M.; J.X.C and A.S provided overall guidance for the project. All authors have given approval to the final version of the manuscript.

**Competing interests:** The authors declare no competing interests.

**Figures and Tables**

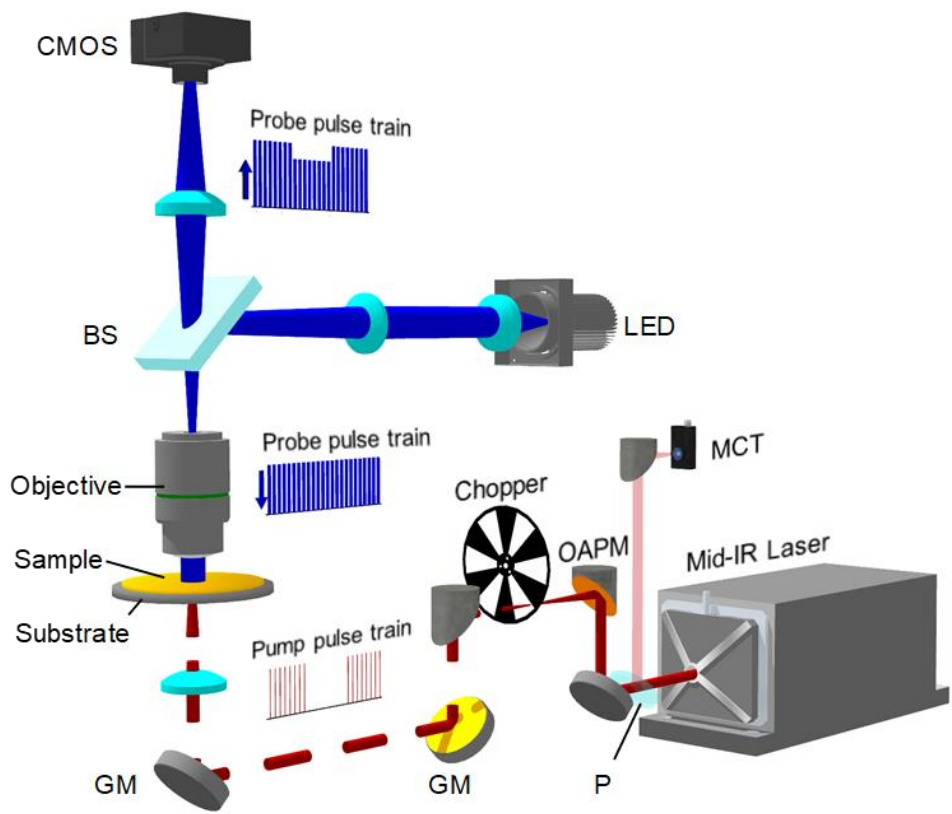

**Fig. 1. Schematic of WPS microscope.** A nanosecond mid-IR laser (bottom right) was sent through an optical chopper and weakly focused on the sample. The IR beam was partially sampled with a $CaF_2$ plate (P) and sent to a mercury cadmium telluride (MCT) detector. The probe was provided by a 450-nm LED, which was imaged to the back aperture of an imaging objective by a 4-f lens system and a 50/50 beam splitter (BS). The sample-reflected light was collected by the same objective and sent to an image sensor with a tube lens. GM, gold mirror; OAPM, off-axis parabolic mirror; LED, light emitting diode; CMOS, complementary metal oxide semiconductor.

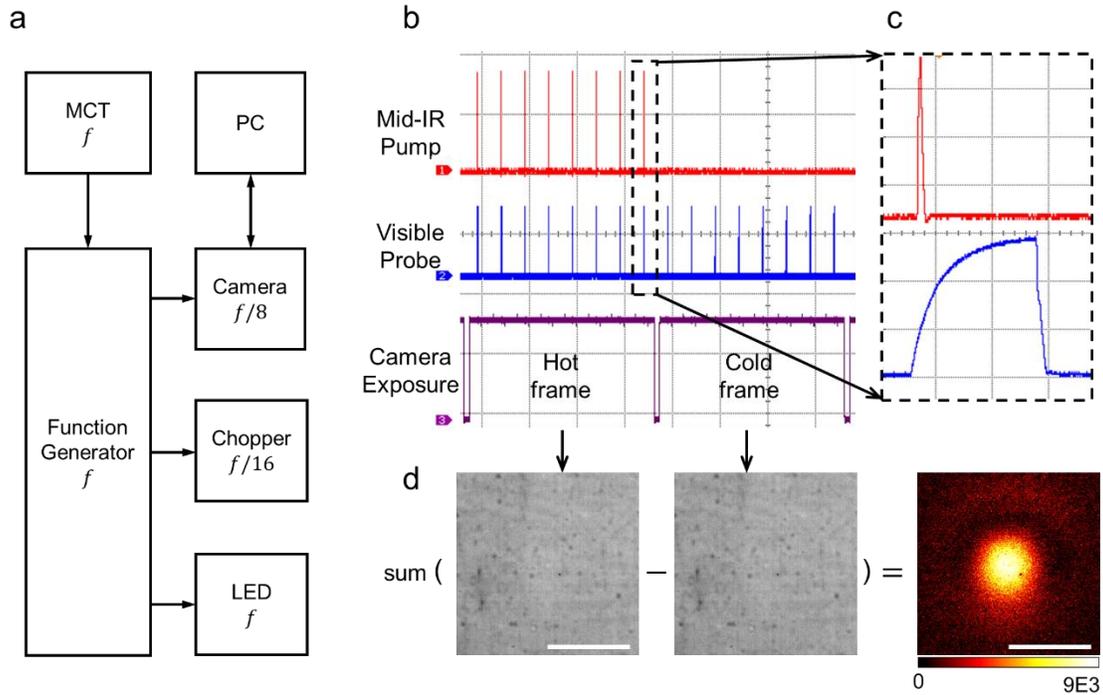

**Fig. 2. Camera-based lock-in detection for WPS imaging.** (a) Block diagram of WPS system. The MCT detector was used to capture the IR laser pulses to generate the master clock $f$ at the repetition rate of the IR laser to trigger the function generator, which sent square wave triggers to the camera, chopper, and LED. The internal frequency divider of the camera was set to expose at $f/8$ frames per second. The chopper divided the trigger pulses by 16 to modulate the pump. A computer was used to control the camera and store the data. (b) Measured pulses of the IR (red), visible (blue), and camera exposure monitor (purple) by an oscilloscope. For the 20 kHz laser repetition rate and 2500 Hz camera framerate, each frame contained eight probe pulses. (c) Zoom-in view of individual pump and probe pulses. The timescale for each grid is 100 μs in (b) and 500 ns in (c). (d) Image processing procedure to generate a WPS image. Contrast was created by subtraction between hot and cold frames. Scale bars, 40 μm.

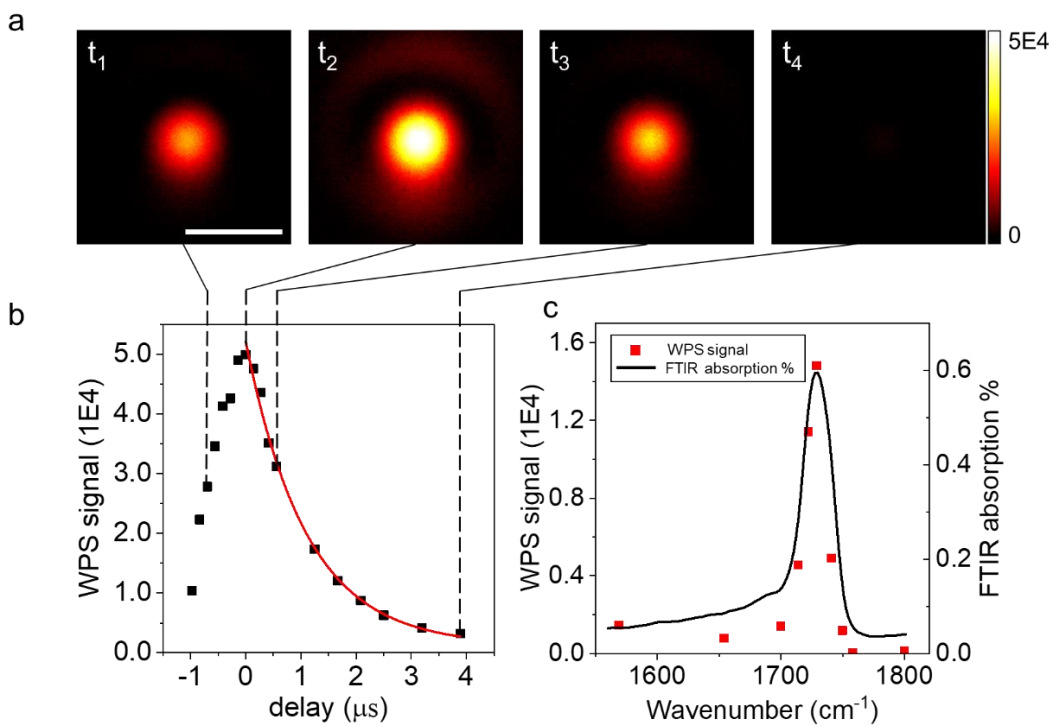

**Fig. 3. Time-resolved WPS imaging of PMMA film on silicon at the 1728 cm$^{-1}$ C=O band.**

(a) WPS images of a 486-nm thick PMMA film at different probe delays. Scale bar: 40 μm. The probe width was 914 ns. Imaging speed: 2 Hz. Imaging contrast: cold − hot. Power at the sample: pump 5.1 mW; probe 1.6 mW. (b) Temporal profile of WPS signal (squares) and the exponential decay fitting result (curve). The decay constant was 1.1 μs. (c) Spectral profile of WPS signal (squares) and the reference FTIR spectrum of PMMA (curve).

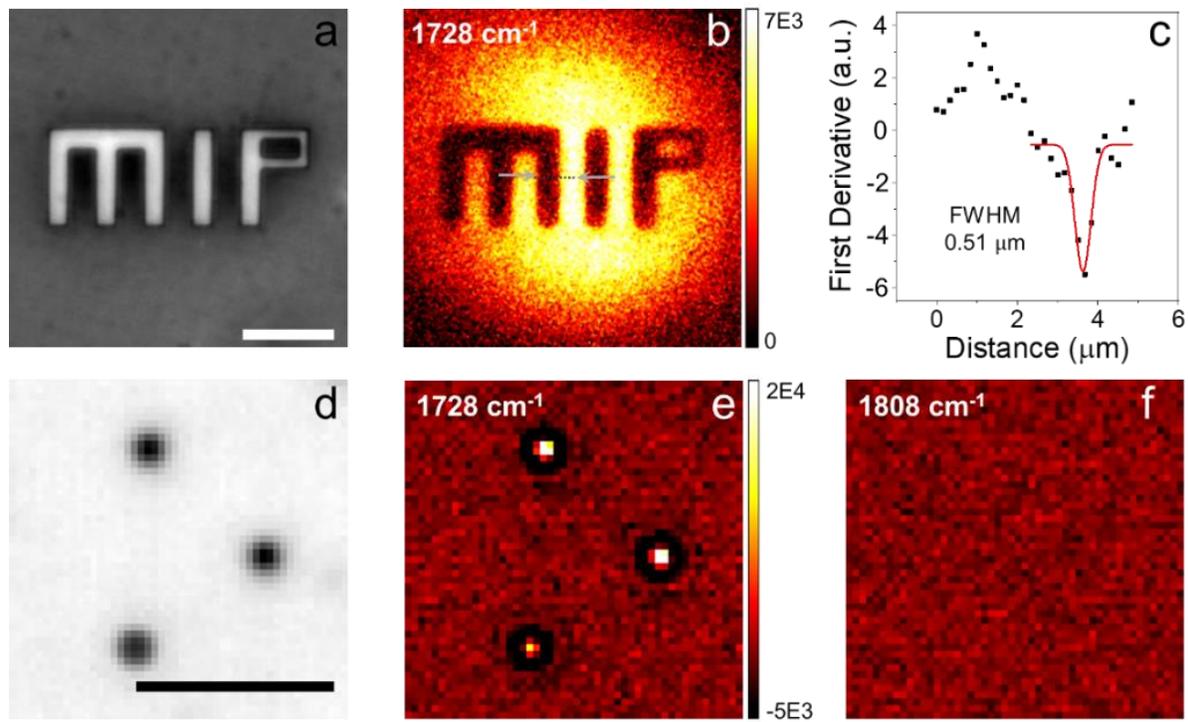

**Fig. 4. WPS imaging of etched pattern in PMMA film.** (a) Reflection image of the pattern, where the etched-off parts showed higher reflectivity. (b) WPS image of the same area. (c) First derivative of the intensity profile along the line shown in (b) as squares. Gaussian fitting (red line) showed an FWHM of 0.51 µm. (d) Reflection image of 1 µm PMMA particles. (e) WPS image of the same area with the pump at 1728 cm$^{-1}$. (f) Off-resonance image showed no contrast. Scale bars: 10 µm.

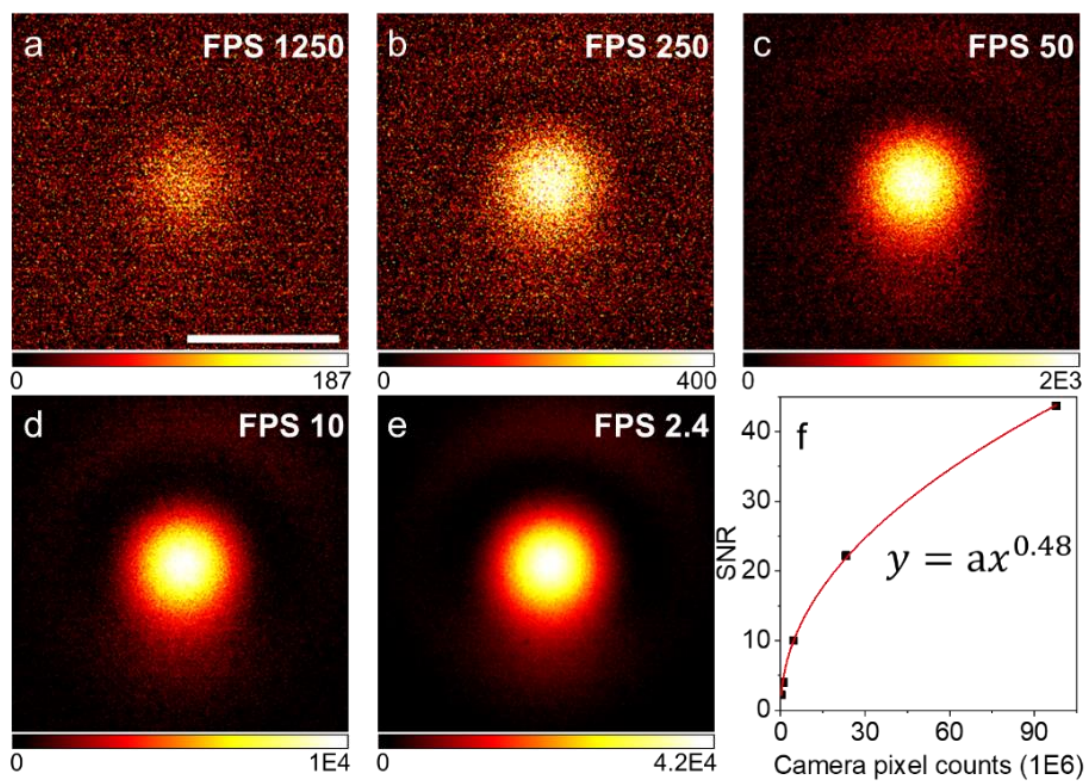

**Fig. 5. Ultrafast chemical imaging of nanoscale PMMA film by WPS microscopy**. (a-e) WPS images of 486-nm thick PMMA film at speed equivalent to 1250, 250, 50, 10, and 2.4 frames per second. (f) Measured SNR and power function fitting result (solid curve). Scale bar, 40 μm.

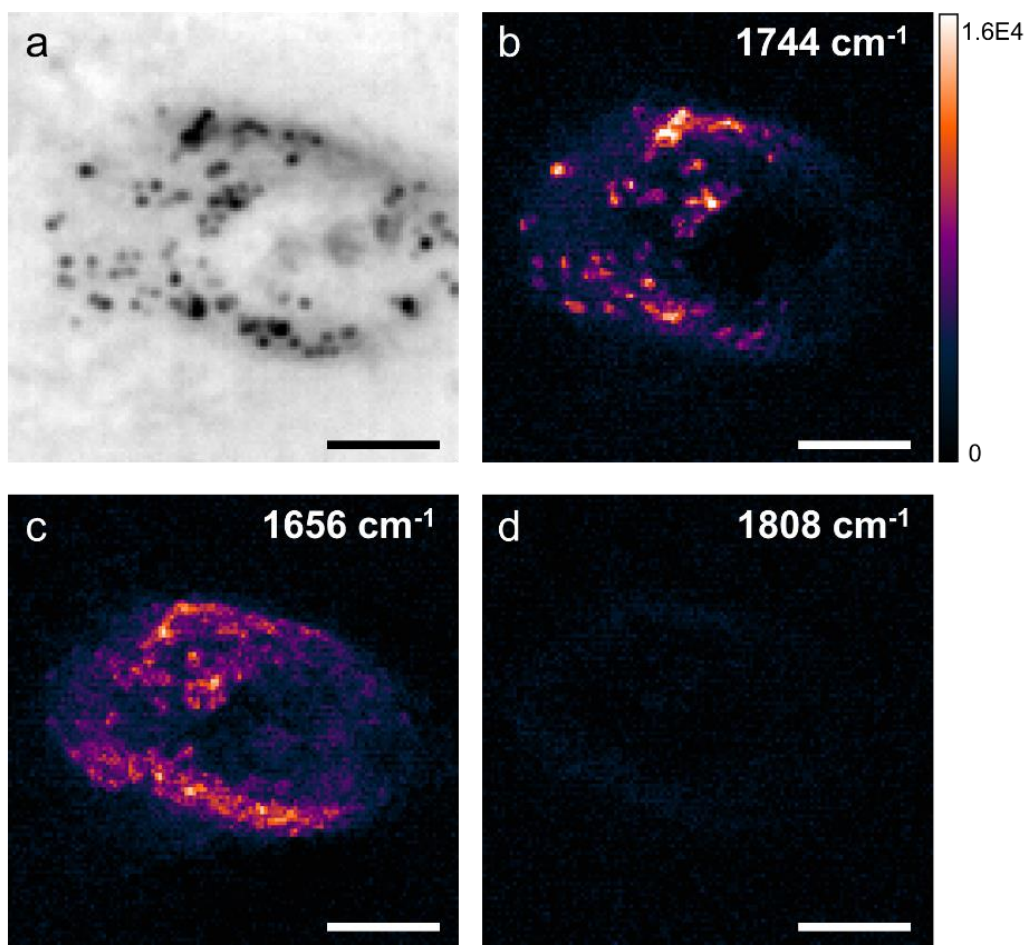

**Fig. 6. WPS imaging of different chemical components in living cells.** (a) Reflection image of a living SKOV3 human ovarian cancer cell cultured on a silicon wafer. (b-d) WPS image of the same field of view at 1744 cm$^{-1}$ (lipid), 1656 cm$^{-1}$ (protein), and 1808 cm$^{-1}$ (off-resonance), respectively. Scale bars, 10 μm.